\documentclass[reprint,superscriptaddress,amsmath,amssymb,pra]{revtex4-2}

\usepackage[utf8]{inputenc}
\usepackage[T1]{fontenc}

\usepackage{graphicx}
\usepackage{hyperref}
\usepackage{placeins}

\usepackage[locale = US]{siunitx} %UK, US, GE, FR, ZA

\hypersetup{
  linktocpage=false,      % no page numbers are clickable
  colorlinks=true,
  linkcolor=blue,
  citecolor=blue,
  urlcolor=blue,
  breaklinks=true,        % break URLs
  pdfborder={0 0 0},      % for removing borders around links
  bookmarksnumbered=true, % If Acrobat bookmarks are requested, include section numbers.
  bookmarksopen=false,    % If Acrobat bookmarks are requested, show them with all the subtrees expanded.
  pdfpagemode={UseOutlines}, % show pdf bookmarks (indices) on startup; does not function all the time
  pdfsubject = {},
  pdfmenubar=true,        % make PDF viewer’s menu bar visible
}

\usepackage			{xcolor}
\definecolor{blue} 		{rgb} {0.25,0.25,0.75}

\newcommand{\diff}{\text{d}}
\DeclareSIUnit{\atomicunits}{a.u.}
\DeclareSIUnit\angstrom{\text {Å}}

\begin{document}

\title{Cross-process interference in single-cycle electron emission from metal needle tips}

\author{Anne Herzig}                \affiliation{Institute of Physics, University of Rostock, D-18059 Rostock, Germany}
\author{Peter Hommelhoff}           \affiliation{Faculty of Physics, Ludwig Maximilian University Munich, D-80799 Munich, and Department of Physics, Friedrich Alexander University Erlangen-Nuremberg, D-91058 Erlangen, Germany}
\author{Eleftherios Goulielmakis}   \affiliation{Institute of Physics, University of Rostock, D-18059 Rostock, Germany}
\author{Thomas Fennel}              \email{thomas.fennel@uni-rostock.de}      \affiliation{Institute of Physics, University of Rostock, D-18059 Rostock, Germany}
\author{Lennart Seiffert}           \email{lennart.seiffert@uni-rostock.de}      \affiliation{Institute of Physics, University of Rostock, D-18059 Rostock, Germany}

\date{\today}

\begin{abstract} % way to long / 600 characters
\noindent Though interference from different emission channels enabled a deeper understanding of strong-field photoemission in atoms and molecules, it remained out of reach for solids. Here, we explore metal needle tips under single-cycle pulses via classical trajectories extended by quantum diffusion and interference and numerical solution of the time-dependent Schrödinger equation. We find interference of direct and backscattered electrons with fringe pattern encoding sub-cycle information on birth times and near-field driven acceleration dynamics, opening routes for ultrafast solid-state metrology.
\end{abstract}

\maketitle

\begin{figure}[t]
    \centering
    \includegraphics[width = \columnwidth]{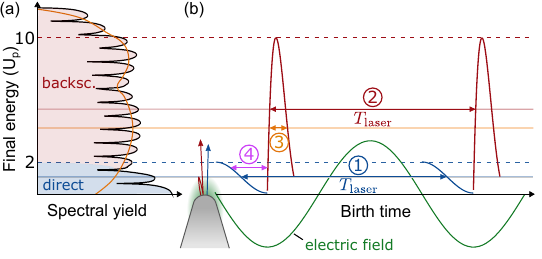}
    \caption{\textbf{Interference mechanisms in strong-field photoemission.} \textbf{(a)}~Schematic sketch of an energy spectrum with typical cut-offs at $\SI{2}{U_\text{p}}$ for direct electrons (blue shaded) and $\SI{10}{U_\text{p}}$ for backscattered electrons (red shaded), ATI peaks and additional modulation due to intracycle interference (orange). \textbf{(b)}~Illustration of birth time-dependent final energies (blue~--~direct, red~--~backscattered). Arrows and lines indicate constructive interference: intercycle (blue and red), intracycle (orange) and possible cross-process (pink) interference.}
    \label{fig:Interferences}
\end{figure}

Strong-field ionization of atoms, molecules as well as nanostructures spotlights the classical and quantum properties of photoemission in at least three ways~\cite{krauszAttosecondPhysics2009, dombiStrongfieldNanooptics2020a, krugerAttosecondPhysicsPhotoemission2012, krugerAttosecondPhysicsPhenomena2018, seiffertStrongfieldPhysicsNanospheres2022a}. First, in a fully classical picture, the trajectories associated with direct and backscattered electrons explain the famous cut-off features at $\SI{2}{U_p}$ and $\SI{10}{U_p}$, respectively~\cite{corkum_plasma_1993, krause_high-order_1992, schafer_above_1993, paulus_rescattering_1994}, see Fig~\ref{fig:Interferences}(a), where $\si{U_p}$ denotes the ponderomotive energy associated with the local optical field. Second, the corresponding quantum trajectories deliver a physical understanding for above-threshold-ionization (ATI) with individual photoelectron peaks separated by the photon energy of the incident laser field (Fig.~\ref{fig:Interferences}(a)). The energy separation $\Delta E_{\rm el}=\Delta(\hbar\omega_{\rm el})=\hbar\omega_{\rm las}$ emerges from constructive interference of the wave packets emitted in subsequent cycles of the laser electric field (intercycle interference) with $\Delta \omega_{\rm el} T_{\rm cycle} =2\pi$, where $T_{\rm cycle}=2\pi/\omega_{\rm las}$, in the sense of a temporal double-slit~\cite{lindnerAttosecondDoubleSlitExperiment2005a, arboTimeDoubleslitInterferences2006}, see labels 1 and 2 in Fig.~\ref{fig:Interferences}(b). Third, additional modulations of ATI spectra on a larger energy scale (orange curve) can be attributed to quantum interference of trajectories originating from a single half-cycle (intracycle interference, cf. label 3 in Fig.~\ref{fig:Interferences}(b))~\cite{arboIntracycleIntercycleInterferences2010a, Shvetsov2018}, highlighting the rich information about  photoemission physics encoded in the spectral features.

Intra- and intercycle interference typically arises from different quantum pathways of the same processes, e.g. self-interference of direct or backscattered electrons. However, also cross-process interference (CPI) between pathways of different processes (cf. label 4 in Fig.~\ref{fig:Interferences}(b)), can lead to characteristic features in photoelectron spectra. For atomic and molecular targets, these include holographic signatures from interference between direct and revisiting electrons~\cite{gopalThreeDimensionalMomentumImaging2009a, khurelbaatarStrongfieldPhotoelectronHolography2024, huismansTimeResolvedHolographyPhotoelectrons2011}, encoding, e.g.,  Coulomb focusing and minute deflection due to the interaction with the residual ion. The visibility of CPI crucially depends on quenching otherwise competing or dominant self-interference effects using intense single-cycle optical waveforms which confine electron emission to only one optical period. The generation of such fields with light-field synthesizers has enabled unprecedented attosecond metrologies applicable to targets ranging from atoms~\cite{hassanOpticalAttosecondPulses2016} over nanosctuctures~\cite{kim_attosecond_2023} to solids~\cite{mouletSoftXrayExcitonics2017}. Whereas CPI from revisiting and backscattered electrons have been identified in $N_2$ and $O_2$~\cite{khurelbaatarStrongfieldPhotoelectronHolography2024} using near single-cycle pulses, the  prerequisites for clean CPI features from direct and backscattered electrons and the resulting prospects are not fully understood.  

\begin{figure}[t]
    \centering
    \includegraphics[width = \columnwidth]{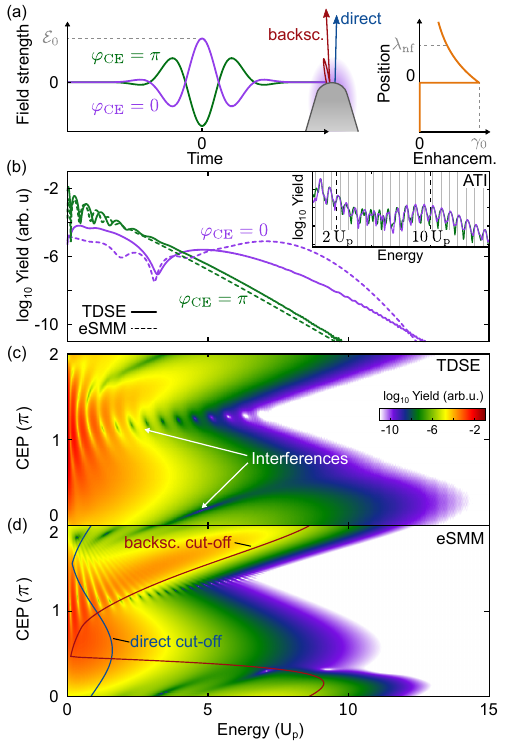}
    \caption{\textbf{CPI in single-cycle driven strong-field electron emission from a metal needle tip.} \textbf{(a)}~Purple and green curves show the incident pulses for two different CEPs (as indicated). Blue and red curves visualize electrons emitted directly and after elastic backscattering. The orange curve shows the spatially decaying near-field profile. \textbf{(b)}~Energy spectra extracted from TDSE (solid) and eSMM (dashed) simulations for the two CEPs as in (a). The inset shows TDSE spectra for long pulses ($12$ cycles) and expected ATI peak positions (grey lines). \textbf{(c,d)}~CEP-dependent energy spectra from TDSE and eSMM simulations. Blue and red curves in (d) indicate the CEP-dependent cut-off energies for direct emission and backscattering.}
    \label{fig:Spec_CEP_TDSE}
\end{figure}

Here we show that CPI between direct and backscattered electrons can be exposed by also suppressing revisiting electrons as achieved for sharp needle tips. To explore the relevant physical mechanism and waveform-dependence, we consider needle tip photoemission under carrier-envelope phase (CEP) controlled single-cycle pulses, cf. Fig.~\ref{fig:Spec_CEP_TDSE}a. Our quantum simulations based on the direct solution of the time-dependent Schrödinger equation for a simplified model system predict strongly CEP-dependent photoelectron spectra containing distinct interference features. Comparison with an extended trajectory model allows the clear association of spectral signatures with direct and backscattered electrons, verifies that the interference features originate from CPI between these processes, and explains the CEP-dependent structures. The interference pattern itself contains information on the relative spectral phase of the direct and backscattered wave packets, the determination of which is central for the temporal and spectral characterization of the ultrashort emitted wave packets~\cite{kim_attosecond_2023}. 

The intrinsic restriction of free electronic motion to only one half-space for needle tips automatically limits emission in time to half-cycles with electric field vectors pointing into the tip and ensures that returning electrons can escape only after backscattering. Despite this seemingly limiting feature and additional modifications due to inhomogeneous near-fields~\cite{herinkFielddrivenPhotoemissionNanostructures2012, heimerlAttosecondPhysicsOptical2025a}, field propagation~\cite{sussmannFieldPropagationinducedDirectionality2015} or charge interactions~\cite{seiffertTrappingFieldAssisted2017,schotzOnsetChargeInteraction2021a},
strong-field photoemission from needle tips (and  nanospheres) resembles the atomic picture to a large extend~\cite{zherebtsovControlledNearfieldEnhanced2011, kruger_attosecond_2011, piglosiewiczCarrierenvelopePhaseEffects2014}. For example, application of the (quantum) trajectory picture to needle tip photoemission uncovered the duration of sub-cycle tunneling~\cite{Dienstbier_N616_2023} and revealed the generation of the shortest electronic signal to date using single-cycle light fields~\cite{kim_attosecond_2023}. 

The quantum description of the needle tip photoemission employed here is adapted from the model described in~\cite{Seiffert2018}. In brief, we solve the one-dimensional time-dependent Schrödinger equation (TDSE) for a single active electron in the spatiotemporal potential $V(x,t) = V_0(x) + V_\text{nf}(x,t)$, where an unperturbed box potential $V_0(x)$ models a metallic tip with work function $W$ and Fermi energy. The time-dependent potential $V_\text{nf}(x,t) = e\int_0^x E_\text{nf}(x',t) \,\diff x'$ reflects the spatio-temporal near-field $E_\text{nf}(x,t)=\gamma(x)E_\text{las}(t)$ resulting from the single-cycle waveform $E_\text{las}(t)$ (purple and green curve in Fig.~\ref{fig:Spec_CEP_TDSE}(a)) and the spatial field enhancement profile 
\begin{equation}
    \gamma(x) = \begin{cases}  1 + (\gamma_0 - 1)\exp(-x/\lambda_\text{nf}) & x \geq 0 \\ 0 & x < 0 \end{cases},
\end{equation}
with peak near-field enhancement $\gamma_0 = 7$ and decay length $\lambda_\text{nf} = \SI{100}{\angstrom}$ as for a typical tungsten needle tip~\cite{thomasLargeOpticalField2015} (orange curve in Fig.~\ref{fig:Spec_CEP_TDSE}(a)). To represent meaningful ultrashort single-cycle waveforms we define the incident electric field $E_\text{las}(t) = -\frac{\partial}{\partial t} A(t)$ via the  vector potential
\begin{equation}
    A(t) = \frac{\mathcal{E}_0}{\omega}\cos(\omega[t-t_0] + \varphi_A)\exp\left(-\frac{(t-t_0)^2}{2\tau^2}\right)
\end{equation}
with peak field strength $\mathcal{E}_0$, central angular frequency $\omega$, pulse duration $\tau$ and the phase of the vector potential $\varphi_A = \varphi_\text{CE} + \pi/2$, where $\varphi_\text{CE}$ is the CEP. This treatment automatically ensures  $\int_{-\infty}^{\infty} E_{\rm las}(t')\,\text{d}t'=0$. We consider single-cycle pulses with $\SI{800}{nm}$  central wavelength, peak intensity $\SI{1e12}{W/cm^2}$ and a duration of $\tau_{_\text{FWHM}} = 2\sqrt{\ln 2} \,\tau=  \pi/\omega \approx \SI{1.3}{fs}$ (intensity FWHM). The resulting peak intensity of the enhanced field at the tip surface of $\SI{4.9e13}{W/cm^2}$ corresponds to a ponderomotive potential of $\si{U_p}=\SI{2.98}{eV}$.

The CEP-dependent photoelectron spectra predicted by the TDSE in Fig.~\ref{fig:Spec_CEP_TDSE}(b,c) exhibit three main features. First, the spectra are strongly modulated by the CEP and do not show clear cut-offs around $\SI{2}{U_p}$ and $\SI{10}{U_p}$ for direct emission and backscattering~\cite{krauszAttosecondPhysics2009} as predicted for long pulses (cf. inset in Fig.~\ref{fig:Spec_CEP_TDSE}(b)). Second, they exhibit unevenly spaced fringe features with a strongly CEP-dependent structure and contrast instead of equidistantly spaced ATI peaks. Third, around $\varphi_\text{CE} = 0$ no discernible fringes but a pronounced dip can be found, which shifts with the CEP. 

To verify that these spectral features reflect CPI of direct and backscattered electrons we performed a semiclassical trajectory analysis~\cite{krause_high-order_1992, schafer_above_1993, corkum_plasma_1993, paulus_rescattering_1994}, extended by quantum diffusion and interference. Trajectories launched at rest for birth times $t_0$ at the classical tunneling exit $x_\text{te} = -W/E_\text{nf}(0,t_0)$ are propagated classically in the local near-field according to
\begin{equation}
    \Ddot{x}(t) = -\frac{e}{m_e}E_\text{nf}(x(t),t).
\end{equation}
For trajectories returning to the surface ($x=0$) we assume elastic backscattering ($\dot x \rightarrow -\dot x$). Each trajectory yields a final energy and momentum $E_\text{f}(t_0)$ and $p_\text{f}(t_0)$ and contributes to the spectrum with a weight $\Gamma(t_0)$ determined from the WKB tunneling rate. This ''simple man's model'' (SMM) predicts CEP-dependent cut-off energies for direct and backscattered electrons (cf. blue and red curves in Fig.~\ref{fig:Spec_CEP_TDSE}(d)) that reveal a phase-dependent switching of the dominance of the individual emission processes: for CEPs around $\varphi_\text{CE} = \SI{0.7}{\pi}$, the direct electrons reach higher energies than the rescattered ones. This represents a remarkable difference to few-cycle or two-color excitation, where the cut-off-modulation typically reflects only backscattered electrons~\cite{kruger_attosecond_2011, Seiffert2018, Dienstbier_N616_2023, piglosiewiczCarrierenvelopePhaseEffects2014}. However, the conventional SMM yields unrealistically sharp spectral cut-off features~\cite{Seiffert2018} and fully neglects interference effects.

\begin{figure}[ht]
    \centering
    \includegraphics[width = 0.95\columnwidth]{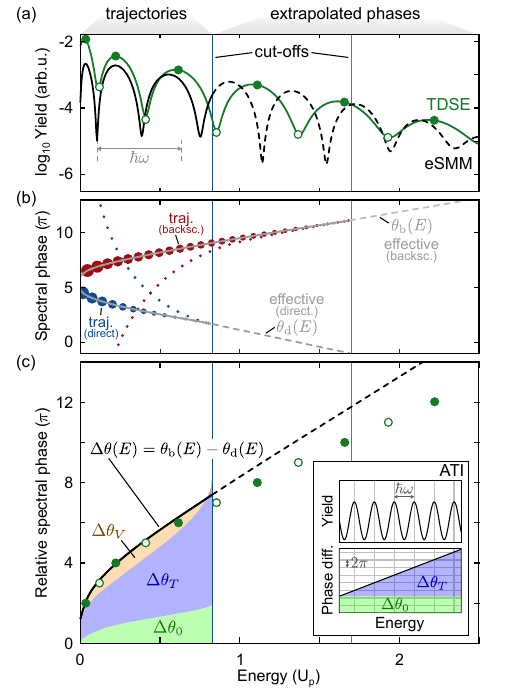}
    \caption{\textbf{Fringe-spacing of CPI and relative spectral phase contributions.}
    \textbf{(a)}~Energy spectra calculated with eSMM (black) and TDSE (green) for $\varphi_\text{CE}=\SI{1.1}{\pi}$. The dashed part indicates the eSMM-spectrum beyond the classical cut-off (cf. vertical blue and red lines for direct and backscattered trajectories) where phases are extrapolated. 
    \textbf{(b)}~Blue and red symbols show final phases and energies of individual direct and backscattered trajectories. The size illustrates the trajectories weights. Curves indicate the final energy-dependent effective phases for the two trajectory classes, which are extrapolated linearly beyond the respective cut-off energies (dashed).
    \textbf{(c)}~Relative spectral phases $\Delta\theta$ between backscattered and direct electrons extracted from the eSMM simulations (black curve) and from the minima and maxima of the TDSE spectrum in (a) (open and closed green symbols). Shaded areas visualize individual contributions of the eSMM-trajectories from the birth time $\Delta\theta_0$ (light green), the kinetic energy $\Delta\theta_T$ (blue) and due to the near-field $\Delta\theta_V$ (orange). The inset shows a respective representation for ATI.}
    \label{fig:analysis_phases}
\end{figure}

\begin{figure*}[ht]
    \centering
    \includegraphics[width = \textwidth]{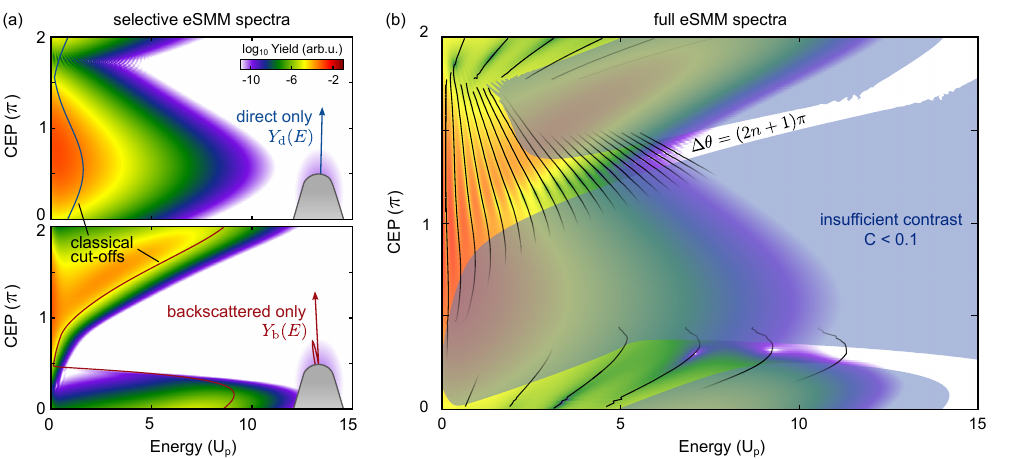}
    \caption{\textbf{Requirements for the emergence of CPI.} \textbf{(a)}~CEP-dependent selective eSMM-spectra $Y_\text{d}(E)$ and $Y_\text{b}(E)$ for direct and backscattered electrons (as indicated). Curves indicate the classical cut-offs as in Fig.~\ref{fig:Spec_CEP_TDSE}d. \textbf{(b)}~CEP-dependent full eSMM-spectra (as in Fig.~\ref{fig:Spec_CEP_TDSE}d) showing CPI when two conditions are fulfilled: 1. High CPI contrast, $C(E)>0.1$, indicated by the non-shaded area. 2. Sufficiently fast energy-evolution of the relative spectral phase $\Delta\theta(E)$. Black curves indicate odd multiples of $\pi$, resulting in destructive interference.}
    \label{fig:eSMM_dir_rec_interf}
\end{figure*}

To extend the description by quantum diffusion and interference~\cite{dienstbierPhDStrongfieldPhysics2022}, we describe each trajectory via the energy-dependent complex amplitude 
\begin{equation}
    \psi(E,t_0)= A(E,t_0) \, e^{i\theta(E)}.
    \label{eq_compamp}
\end{equation}
Here $A(E,t_0)$ describes the spectral broadening according to a Gaussian wave packet in momentum space, expressed in terms of energy via
\begin{equation}
    A(E,t_0) = c\, \sqrt{\Gamma(t_0)} \, e^{-\frac{\left( \sqrt{E_\text{f}(t_0)} - \sqrt{E} \right)^2}{\sigma^2}},
\end{equation}
with normalization constant $c$ and spectral width $\sigma^2$. The effective phase function $\theta(E)$ in Eq.~\eqref{eq_compamp} is constructed in WKB approximation from the set of birth times for which trajectories match the considered energy $E$ via phase-correct accumulation of respective amplitudes according to 
\begin{equation}
\theta(E)=\text{arg}\!\left[\int\limits_{-\infty}^t\!\sqrt{\Gamma(t_0)} e^{\frac{i}{\hbar}S_\text{c}(t_0, t)} \delta(E\!-\!E_\text{f}(t_0)) \,\text{d}t_0 \right]\!.
\end{equation}
In the compensated action 
\begin{equation}
 S_\text{c}(t_0,t) = S(t_0,t) - p_\text{f}(t_0)x(t) + E_\text{f}(t_0) t    \label{eq:compensation}
\end{equation}
the trivial time dependence associated with the asymptotic drift motion is removed such that 
the effective phase becomes stationary for sufficiently late evaluation time $t$.
The trajectory-specific classical action required to evaluate the effective phase reads~\cite{Goldfarb_JCP128_2008}
\begin{equation}
    S(t_0,t)=Wt_\text{0} + \int\limits_{t_\text{0}}^t \frac{m_e}{2}v^2(t')\text{d}t'-\int\limits_{t_\text{0}}^t  V_{\rm nf}(x(t'),t') \text{d}t'. \label{eq:action}
\end{equation}

The three terms on the right hand side reflect the specific phase contributions $\theta_0$, $\theta_T$, and $\theta_V$ encoding the birth time, the kinetic energy evolution, and the near-field effect, respectively. Here, we always include the phase compensation resulting from Eq.~\ref{eq:compensation} in the kinetic energy contribution. We emphasize that the near-field phase contribution is neglected within the commonly employed strong-field approximation that assumes a homogeneous field but is crucial for our scenario. Note that a trajectory-selective analysis of the phase can yield multiple contributions for a given energy (cf. colored symbols in Fig.~\ref{fig:analysis_phases}(b)) while the effective phase is dominated by the trajectories with particularly high weight. Selective application of the above procedure to direct and backscattered electrons yields the corresponding effective phase functions $\theta_\text{d}(E)$ and $\theta_\text{b}(E)$ up to the specific cut-off energies (solid curves in Fig.~\ref{fig:analysis_phases}(b)) and are linearly extrapolated beyond the cut-offs (dashed curves in Fig.~\ref{fig:analysis_phases}(b)). Note that an additional phase shift of $\pi$ is included upon backscattering. 

Within this ''extended simple man's model'' (eSMM), energy spectra  follow from integration of the process-selective complex amplitudes $\psi_\text{d}(E,t_0)$ and $\psi_\text{b}(E,t_0)$ over all birth times via \begin{equation}
    Y(E) = \left| \int \left(\psi_\text{d}(E, t_0)+\psi_\text{b}(E, t_0)\right) \, \text{d}t_0\right|^2.
\end{equation}
For our scenario we used spectral widths $\sigma^2_\text{d} = \SI{1.97}{\eV}$ and $\sigma^2_\text{b} = \SI{0,36}{eV}$ in the spectral broadening for best agreement with the TDSE results. Though our approximation does not enforce norm conservation strictly, which would require an advanced description by, e.g., wave packet molecular dynamics~\cite{grabowskiReviewWavePacket2014}, our simplified model shows remarkable agreement with TDSE (compare Fig.~\ref{fig:Spec_CEP_TDSE}(b,d) to Fig.~\ref{fig:Spec_CEP_TDSE}(b,c)). In particular, the CEP-dependent modulation of the spectra as well as the pronounced interference structures are reproduced, supporting that eSMM captures the central physical processes and is suited for their further detailed analysis.

Regarding the pronounced interference features, the eSMM unambiguously reveals that the fringes result from CPI of direct and backscattered electrons, because the interference vanishes in the individual eSMM spectra (cf. $Y_\text{d}(E)$ and $Y_\text{b}(E)$ in Fig.~\ref{fig:eSMM_dir_rec_interf}(a)). Note that only minor residual self-interference signals remain in the direct electrons ($\varphi_\text{CE}\approx 1.8\pi$) and in the backscattered electrons ($\varphi_\text{CE}\approx 0.2\pi$). The selective spectra further explain that the CPI contrast 
\begin{equation}
    C(E) = \frac{2 \sqrt{Y_\text{d}(E)} \sqrt{Y_\text{b}(E)}}{Y_\text{d}(E) + Y_\text{b}(E)}
\end{equation}
is responsible for the emergence of CPI fringes only in specific regions of the CEP-energy map as highlighted via a contrast mask in Fig.~\ref{fig:eSMM_dir_rec_interf}(b).
An additional prerequisite is a sufficiently fast evolution of the relative spectral phase $\Delta\theta(E) = \theta_\text{b}(E) - \theta_\text{d}(E)$ of direct and backscattered electrons with energy. In this case, constructive and destructive interference (cf. black curves in Fig.~\ref{fig:eSMM_dir_rec_interf}(b)) alternate rapidly as function of energy, leading to fringes. CPI hence only becomes visible in regions where sufficiently high contrast and rapidly evolving relative spectral phase occur.

Having identified the prerequisites for CPI fringes, we now explore the physics encoded in their particular shape, namely their energy-dependent position and spacing. The representative example in Fig.~\ref{fig:analysis_phases}(a) shows that the spacing does not match the photon energy (grey arrow) and that the eSMM prediction evolves very similarly to the TDSE result up to the classical cut-off of direct electrons (compare solid black to green curve). The larger deviation at higher energies is attributed to the uncertainties resulting from the required phase extrapolation. The relative spectral phase evolution from eSMM (black curve in Fig.~\ref{fig:analysis_phases}(c)) can further be decomposed into the individual contributions from birth time, kinetic energy and near-field effect (shaded areas). For comparison, we also extract the relative spectral phase evolution from the position of minima and maxima in the TDSE spectrum (open and closed green symbols in Figs.~\ref{fig:analysis_phases}(a,c)). 

Three central conclusions can be drawn from this comparison. First, the eSMM result matches the evolution of the TDSE result remarkably well up to the direct electron cut-off and also the extrapolation shows the same trend, i.e., a linear evolution with energy beyond the cut-off, albeit with a slightly different slope. Second, focusing on the region below the cut-off shows that the slope of the relative spectral phase decreases with energy, explaining the increasing spacing of the fringes. Third and most importantly, it is imperative to take all three relative spectral phase contributions into account, in particular the near-field effect, which is commonly completely neglected. This is in fundamental contrast to the case of ATI in longer pulses, where the near-field effect of trajectories from adjacent cycles perfectly cancels and the strictly linear evolution of the kinetic energy term with energy creates equally spaced peaks, separated by the photon energy (see inset in Fig.~\ref{fig:analysis_phases}(c)). The fact that all three contributions, i.e., the birth time, the  kinetic energy and the near-field effect, enter the relative spectral phase evolution in a nontrivial way opens routes for their selective investigation via the CPI fringe structure. 

In conclusion, our theoretical analysis of single-cycle driven photoemission from sharp needle tips predicts robust CEP-dependent interference structures caused by cross-process interference (CPI) of direct and backscattered electrons. By comparison of TDSE results with an extended simple man's model, we show under which conditions CPI occurs and how the nontrivial fringe pattern can be explained with the relative spectral phase evolution  of direct and backscattered electrons. Our findings highlight that CPI fringes reveal a distinct near-field contribution that extends beyond the established quantum trajectory description within strong-field approximation. The CPI features from direct and backscattered electrons encode rich information on the near-field driven acceleration dynamics, in principle allowing its characterization via the corresponding spectra. Our work provides the theoretical toolbox to analyze such phenomena and to develop corresponding reconstruction approaches in the future.

\subsection*{Acknowledgments}
We thank Philip Dienstbier for fruitful discussions. We acknowledge financial support from the German Research Foundation (DFG) via CRC 1477 ”Light-Matter Interactions at Interfaces” (ID: 441234705).

\bibliographystyle{style_new}
\bibliography{references}

\begin{thebibliography}{10}

\bibitem[1]{krauszAttosecondPhysics2009}
F.~Krausz and M.~Ivanov,
\newblock {\em Attosecond physics},
\newblock \href{http://dx.doi.org/10.1103/RevModPhys.81.163}{Reviews of Modern
  Physics \textbf{81}, 163--234 (2009)}.

\bibitem[2]{dombiStrongfieldNanooptics2020a}
P.~Dombi, Z.~Pápa, J.~Vogelsang, S.~V. Yalunin, M.~Sivis, G.~Herink,
  S.~Schäfer, P.~Groß, C.~Ropers and C.~Lienau,
\newblock {\em Strong-field nano-optics},
\newblock \href{http://dx.doi.org/10.1103/RevModPhys.92.025003}{Reviews of
  Modern Physics \textbf{92}, 025003 (2020)}.

\bibitem[3]{krugerAttosecondPhysicsPhotoemission2012}
M.~Krüger, M.~Schenk, M.~Förster and P.~Hommelhoff,
\newblock {\em Attosecond physics in photoemission from a metal nanotip},
\newblock \href{http://dx.doi.org/10.1088/0953-4075/45/7/074006}{Journal of
  Physics B: Atomic, Molecular and Optical Physics \textbf{45}, 074006 (2012)}.

\bibitem[4]{krugerAttosecondPhysicsPhenomena2018}
M.~Krüger, C.~Lemell, G.~Wachter, J.~Burgdörfer and P.~Hommelhoff,
\newblock {\em Attosecond physics phenomena at nanometric tips},
\newblock \href{http://dx.doi.org/10.1088/1361-6455/aac6ac}{Journal of Physics
  B: Atomic, Molecular and Optical Physics \textbf{51}, 172001 (2018)}.

\bibitem[5]{seiffertStrongfieldPhysicsNanospheres2022a}
L.~Seiffert, S.~Zherebtsov, M.~F. Kling and T.~Fennel,
\newblock {\em Strong-field physics with nanospheres},
\newblock \href{http://dx.doi.org/10.1080/23746149.2021.2010595}{Advances in
  Physics: X \textbf{7}, 2010595 (2022)}.

\bibitem[6]{corkum_plasma_1993}
P.~B. Corkum,
\newblock {\em Plasma perspective on strong field multiphoton ionization},
\newblock \href{http://dx.doi.org/10.1103/PhysRevLett.71.1994}{Physical Review
  Letters \textbf{71}, 1994--1997 (1993)}.

\bibitem[7]{krause_high-order_1992}
J.~L. Krause, K.~J. Schafer and K.~C. Kulander,
\newblock {\em High-order harmonic generation from atoms and ions in the high
  intensity regime},
\newblock \href{http://dx.doi.org/10.1103/PhysRevLett.68.3535}{Physical Review
  Letters \textbf{68}, 3535--3538 (1992)}.

\bibitem[8]{schafer_above_1993}
K.~J. Schafer, B.~Yang, L.~F. DiMauro and K.~C. Kulander,
\newblock {\em Above threshold ionization beyond the high harmonic cutoff},
\newblock \href{http://dx.doi.org/10.1103/PhysRevLett.70.1599}{Physical Review
  Letters \textbf{70}, 1599--1602 (1993)}.

\bibitem[9]{paulus_rescattering_1994}
G.~G. Paulus, W.~Becker, W.~Nicklich and H.~Walther,
\newblock {\em Rescattering effects in above-threshold ionization: a classical
  model},
\newblock \href{http://dx.doi.org/10.1088/0953-4075/27/21/003}{Journal of
  Physics B: Atomic, Molecular and Optical Physics \textbf{27}, L703 (1994)}.

\bibitem[10]{lindnerAttosecondDoubleSlitExperiment2005a}
F.~Lindner, M.~G. Schätzel, H.~Walther, A.~Baltuška, E.~Goulielmakis,
  F.~Krausz, D.~B. Milošević, D.~Bauer, W.~Becker and G.~G. Paulus,
\newblock {\em Attosecond {Double}-{Slit} {Experiment}},
\newblock \href{http://dx.doi.org/10.1103/PhysRevLett.95.040401}{Physical
  Review Letters \textbf{95}, 040401 (2005)}.

\bibitem[11]{arboTimeDoubleslitInterferences2006}
D.~G. Arbó, E.~Persson and J.~Burgdörfer,
\newblock {\em Time double-slit interferences in strong-field tunneling
  ionization},
\newblock \href{http://dx.doi.org/10.1103/PhysRevA.74.063407}{Physical Review A
  \textbf{74}, 063407 (2006)}.

\bibitem[12]{arboIntracycleIntercycleInterferences2010a}
D.~G. Arbó, K.~L. Ishikawa, K.~Schiessl, E.~Persson and J.~Burgdörfer,
\newblock {\em Intracycle and intercycle interferences in above-threshold
  ionization: {The} time grating},
\newblock \href{http://dx.doi.org/10.1103/PhysRevA.81.021403}{Physical Review A
  \textbf{81}, 021403 (2010)}.

\bibitem[13]{Shvetsov2018}
N.~I. Shvetsov-Shilovski and M.~Lein,
\newblock {\em Effects of the Coulomb potential in interference patterns of
  strong-field holography with photoelectrons},
\newblock \href{http://dx.doi.org/10.1103/PhysRevA.97.013411}{Phys. Rev. A
  \textbf{97}, 013411 (2018)}.

\bibitem[14]{gopalThreeDimensionalMomentumImaging2009a}
R.~Gopal, K.~Simeonidis, R.~Moshammer, T.~Ergler, M.~Dürr, M.~Kurka, K.-U.
  Kühnel, S.~Tschuch, C.-D. Schröter, D.~Bauer, J.~Ullrich, A.~Rudenko,
  O.~Herrwerth, T.~Uphues, M.~Schultze, E.~Goulielmakis, M.~Uiberacker,
  M.~Lezius and M.~F. Kling,
\newblock {\em Three-{Dimensional} {Momentum} {Imaging} of {Electron} {Wave}
  {Packet} {Interference} in {Few}-{Cycle} {Laser} {Pulses}},
\newblock \href{http://dx.doi.org/10.1103/PhysRevLett.103.053001}{Physical
  Review Letters \textbf{103}, 053001 (2009)}.

\bibitem[15]{khurelbaatarStrongfieldPhotoelectronHolography2024}
T.~Khurelbaatar, J.~Heo, S.~Yu, X.~Lai, X.~Liu and D.~E. Kim,
\newblock {\em Strong-field photoelectron holography in the subcycle limit},
\newblock \href{http://dx.doi.org/10.1038/s41377-024-01457-7}{Light: Science \&
  Applications \textbf{13}, 108 (2024)}.

\bibitem[16]{huismansTimeResolvedHolographyPhotoelectrons2011}
Y.~Huismans, A.~Rouzée, A.~Gijsbertsen, J.~H. Jungmann, A.~S. Smolkowska,
  P.~S. W.~M. Logman, F.~Lépine, C.~Cauchy, S.~Zamith, T.~Marchenko, J.~M.
  Bakker, G.~Berden, B.~Redlich, A.~F.~G. van~der Meer, H.~G. Muller,
  W.~Vermin, K.~J. Schafer, M.~Spanner, M.~Y. Ivanov, O.~Smirnova, D.~Bauer,
  S.~V. Popruzhenko and M.~J.~J. Vrakking,
\newblock {\em Time-{Resolved} {Holography} with {Photoelectrons}},
\newblock \href{http://dx.doi.org/10.1126/science.1198450}{Science
  \textbf{331}, 61--64 (2011)}.

\bibitem[17]{hassanOpticalAttosecondPulses2016}
M.~T. Hassan, T.~T. Luu, A.~Moulet, O.~Raskazovskaya, P.~Zhokhov, M.~Garg,
  N.~Karpowicz, A.~M. Zheltikov, V.~Pervak, F.~Krausz and E.~Goulielmakis,
\newblock {\em Optical attosecond pulses and tracking the nonlinear response of
  bound electrons},
\newblock \href{http://dx.doi.org/10.1038/nature16528}{Nature \textbf{530},
  66--70 (2016)}.

\bibitem[18]{kim_attosecond_2023}
H.~Y. Kim, M.~Garg, S.~Mandal, L.~Seiffert, T.~Fennel and E.~Goulielmakis,
\newblock {\em Attosecond field emission},
\newblock \href{http://dx.doi.org/10.1038/s41586-022-05577-1}{Nature
  \textbf{613}, 662--666 (2023)}.

\bibitem[19]{mouletSoftXrayExcitonics2017}
A.~Moulet, J.~B. Bertrand, T.~Klostermann, A.~Guggenmos, N.~Karpowicz and
  E.~Goulielmakis,
\newblock {\em Soft x-ray excitonics},
\newblock \href{http://dx.doi.org/10.1126/science.aan4737}{Science
  \textbf{357}, 1134--1138 (2017)}.

\bibitem[20]{herinkFielddrivenPhotoemissionNanostructures2012}
G.~Herink, D.~R. Solli, M.~Gulde and C.~Ropers,
\newblock {\em Field-driven photoemission from nanostructures quenches the
  quiver motion},
\newblock \href{http://dx.doi.org/10.1038/nature10878}{Nature \textbf{483},
  190--193 (2012)}.

\bibitem[21]{heimerlAttosecondPhysicsOptical2025a}
J.~Heimerl, S.~Meier, A.~Herzig, F.~L. Hoffmann, L.~Seiffert, D.~Lesko,
  S.~Hillmann, S.~Wittigschlager, T.~Weitz, T.~Fennel and P.~Hommelhoff,
\newblock {\em Attosecond physics in optical near fields},
\newblock \href{http://dx.doi.org/10.48550/arXiv.2507.02673}{ ,  (2025)}.

\bibitem[22]{sussmannFieldPropagationinducedDirectionality2015}
F.~Süßmann, L.~Seiffert, S.~Zherebtsov, V.~Mondes, J.~Stierle, M.~Arbeiter,
  J.~Plenge, P.~Rupp, C.~Peltz, A.~Kessel, S.~A. Trushin, B.~Ahn, D.~Kim,
  C.~Graf, E.~Rühl, M.~F. Kling and T.~Fennel,
\newblock {\em Field propagation-induced directionality of carrier-envelope
  phase-controlled photoemission from nanospheres},
\newblock \href{http://dx.doi.org/10.1038/ncomms8944}{Nature Communications
  \textbf{6}, 7944 (2015)}.

\bibitem[23]{seiffertTrappingFieldAssisted2017}
L.~Seiffert, P.~Henning, P.~Rupp, S.~Zherebtsov, P.~Hommelhoff, M.~F. Kling and
  T.~Fennel,
\newblock {\em Trapping field assisted backscattering in strong-field
  photoemission from dielectric nanospheres},
\newblock \href{http://dx.doi.org/10.1080/09500340.2017.1288838}{Journal of
  Modern Optics \textbf{64}, 1096--1103 (2017)}.

\bibitem[24]{schotzOnsetChargeInteraction2021a}
J.~Schötz, L.~Seiffert, A.~Maliakkal, J.~Blöchl, D.~Zimin, P.~Rosenberger,
  B.~Bergues, P.~Hommelhoff, F.~Krausz, T.~Fennel and M.~F. Kling,
\newblock {\em Onset of charge interaction in strong-field photoemission from
  nanometric needle tips},
\newblock \href{http://dx.doi.org/10.1515/nanoph-2021-0276}{Nanophotonics
  \textbf{10}, 3769--3775 (2021)}.

\bibitem[25]{zherebtsovControlledNearfieldEnhanced2011}
S.~Zherebtsov, T.~Fennel, J.~Plenge, E.~Antonsson, I.~Znakovskaya, A.~Wirth,
  O.~Herrwerth, F.~Süßmann, C.~Peltz, I.~Ahmad, S.~A. Trushin, V.~Pervak,
  S.~Karsch, M.~J.~J. Vrakking, B.~Langer, C.~Graf, M.~I. Stockman, F.~Krausz,
  E.~Rühl and M.~F. Kling,
\newblock {\em Controlled near-field enhanced electron acceleration from
  dielectric nanospheres with intense few-cycle laser fields},
\newblock \href{http://dx.doi.org/10.1038/nphys1983}{Nature Physics \textbf{7},
  656--662 (2011)}.

\bibitem[26]{kruger_attosecond_2011}
M.~Krüger, M.~Schenk and P.~Hommelhoff,
\newblock {\em Attosecond control of electrons emitted from a nanoscale metal
  tip},
\newblock \href{http://dx.doi.org/10.1038/nature10196}{Nature \textbf{475},
  78--81 (2011)}.

\bibitem[27]{piglosiewiczCarrierenvelopePhaseEffects2014}
B.~Piglosiewicz, S.~Schmidt, D.~J. Park, J.~Vogelsang, P.~Groß, C.~Manzoni,
  P.~Farinello, G.~Cerullo and C.~Lienau,
\newblock {\em Carrier-envelope phase effects on the strong-field photoemission
  of electrons from metallic nanostructures},
\newblock \href{http://dx.doi.org/10.1038/nphoton.2013.288}{Nature Photonics
  \textbf{8}, 37--42 (2014)}.

\bibitem[28]{Dienstbier_N616_2023}
P.~Dienstbier, L.~Seiffert, T.~Paschen, A.~Liehl, A.~Leitenstorfer, T.~Fennel
  and P.~Hommelhoff,
\newblock {\em Tracing attosecond electron emission from a nanometric metal
  tip},
\newblock \href{http://dx.doi.org/10.1038/s41586-023-05839-6}{Nature
  \textbf{616}, 702--706 (2023)}.

\bibitem[29]{Seiffert2018}
L.~Seiffert, T.~Paschen, P.~Hommelhoff and T.~Fennel,
\newblock {\em High-order above-threshold photoemission from nanotips
  controlled with two-color laser fields},
\newblock \href{http://dx.doi.org/10.1088/1361-6455/aac34f}{Journal of Physics
  B: Atomic, Molecular and Optical Physics \textbf{51},  (2018)}.

\bibitem[30]{thomasLargeOpticalField2015}
S.~Thomas, G.~Wachter, C.~Lemell, J.~Burgdörfer and P.~Hommelhoff,
\newblock {\em Large optical field enhancement for nanotips with large opening
  angles},
\newblock \href{http://dx.doi.org/10.1088/1367-2630/17/6/063010}{New Journal of
  Physics \textbf{17}, 063010 (2015)}.

\bibitem[31]{dienstbierPhDStrongfieldPhysics2022}
P.~Dienstbier,
\newblock {\em {Strong}-field physics at nanoemitters in tailored complex
  light-fields},
\newblock PhD thesis,
\newblock
  \href{https://open.fau.de/items/ec6f4380-611b-48aa-9db4-81167d2a13ed}{
  (2022)}.

\bibitem[32]{Goldfarb_JCP128_2008}
Y.~Goldfarb, J.~Schiff and D.~J. Tannor,
\newblock {\em Complex trajectory method in time-dependent WKB},
\newblock \href{http://dx.doi.org/10.1063/1.2907336}{The Journal of Chemical
  Physics \textbf{128}, 164114 (2008)}.

\bibitem[33]{grabowskiReviewWavePacket2014}
P.~E. Grabowski,
\newblock {\em A {Review} of {Wave} {Packet} {Molecular} {Dynamics}},
\newblock \href{http://dx.doi.org/10.1007/978-3-319-04912-0_10}{ , 265--282
  (2014)}.

\end{thebibliography}

\end{document}